# DESIGN OF A MARX-TOPOLOGY MODULATOR FOR FNAL LINAC*

T.A. Butler#, F.G. Garcia, M.R. Kufer, H. Pfeffer, D. Wolff, FNAL, Batavia, IL 60510, USA

## Abstract

The Fermilab Proton Improvement Plan (PIP) was formed in late 2011 to address important and necessary upgrades to the Proton Source machines (Injector line, Linac and Booster). The goal is to increase the proton flux by doubling the Booster beam cycle rate while maintaining the same intensity per cycle, the same uptime, and the same residual activation in the enclosure. For the Linac, the main focus within PIP is to address reliability. One of the main tasks is to replace the present hard-tube modulator used on the 200 MHz RF system. Plans to replace this high power system with a Marx-topology modulator, capable of providing the required waveform shaping to stabilize the accelerating gradient and compensate for beam loading, will be presented, along with development data from the prototype unit.

## INTRODUCTION

The Fermi National Accelerator Laboratory (FNAL) Linear accelerator accelerates H- beam pulses from 750 keV up to 400 MeV. The lower energy Linac section, built in 1969, uses 201.25 MHz Alvarez style drift tube cavities powered by a 5 MW triode power tube (therein referred as 7835) accelerating beam up to 116 MeV, while the higher energy section uses 805 MHz side coupled cavities powered by a klystron to boost the beam energy to 400 MeV. The 7835 is plate modulated using a hard-tube topology modulator where a high voltage capacitor bank provides energy that is switched via three parallel grid-controlled electron tubes to regulate the anode on the triode. These series-pass tubes were discontinued in the early 2000s, so in order to sustain operations, these tubes required the development and management of a dedicated maintenance schedule in order to rebuild depleted tubes. This approach has been extending the life of the present modulator, but is not sustainable for long term operations.

Within the Proton Improvement Plan (PIP) [1], the chosen path to mitigate this reliability issue was to replace the hard-tube modulator with a modern solid state modulator which improves reliability, lowers operational costs while maintaining the same waveform accuracy required to accelerate the beam. The main challenge in designing a replacement for the present hard-tube modulator is achieving the waveform shaping that is required to regulate the accelerating cavity fields. Unlike other Linac RF systems, which typically use direct feedback (driving the RF input to the power amplifier to regulate gradient), the 7835 triode amplifier uses the plate voltage to regulate the cavity field, requiring a modulator capable of creating an adaptable waveform to compensate for beam loading and overshoot for filling the cavity.

## PRESENT MODULATOR TOPOLOGY AND SPECIFICATIONS

The present tube based modulator acts essentially as a large operational amplifier with finite gain. When operated in an analog feedback loop, it has the ability to regulate its output to any desired shape, which is presently chosen as a trapezoidal ramped waveform. Furthermore, this system compensates in real time for pulse-to-pulse gain changes, which are present in many amplifier stages. By choosing the feedback signal as the peak detected accelerating cavity gradient, the feedback system is able to regulate the cavity fields with ~0.2%. In Table 1, a subset of the most critical parameters for the modulator are listed.

Table 1: Linac Modulator Specifications

| Parameter | Value | Units |
|---|---|---|
| Maximum Voltage | 35 | kV |
| Maximum Current | 375 | A |
| Voltage Regulation | ±25 | V |
| Minimum Slew Rate | 15 | kV/µs |
| Maximum Rise/Fall Step | 1.5 | kV |
| Maximum Beam Step | 8 | kV |
| Maximum Beam Tilt | ±5 | kV |
| Maximum Pulse Width | 460 | µs |
| Variable Rise/Fall Time | 50-150 | µs |
| Pulse Repetition Rate | 15 | Hz |

## NEW MODULATOR DESIGN

A careful study of the numerous alternative options were conducted and down-selected to the two best designs. Among the alternative options were 1) replace the modulator with another hard-tube modulator; 2) replace with industry developed solid state modulator and 3) adapt the Marx-topology modulator developed by SLAC for the International Linear Collider (ILC) project [2]. Despite having strong features most of the options would involve significant development costs in order to meet the desired operational requirements. In particular, the later was extensively studied, but it had slightly higher cost and more engineering time required to convert the design to operate in real time feedback, which is critical for meeting the desired operational requirements [3]. Therefore, the best option chosen for Linac was to design a new system, based on Marx-topology modulator which was designed to meet the critical specifications with the ability to run in real time feedback without compromising any other requirements.

The Fermilab Accelerator Division Electrical Engineering Department (AD/EE) took the lead on this development.

Undoubtedly, the highly important and challenging specifications, such as ± 5 kV beam top tilt; 15 kV/µs slew rate; ± 25 V flattop beam regulation; maximum spark energy of 5 J, and a maximum step size of 1.5 kV were all critical in determining the ideal cell size, voltage per cell, di/dt limitations and Marx cell topology. Since the design requires a fast slew rate to compensate for beam loading and a stringent ripple requirement, the Marx cells were divided into two main groups: i) switching cells which are used to create the basic waveform and produce the fast beam pulse rise/fall time and ii) regulating cells which are used to compensate for capacitor droop and provide the necessary beam top tilt. The full voltage modulator is made up of 54 cells which will be discharged in series, in specific quantities at specific times, to create the desired waveform. These cells are grouped as 12 regulating cells, 41 switching cells, and one beam step cell. The switching and regulating cells will be driven from a 1 kV charging supply set to approximately 925 V for operation. A simplified schematic of the modulator is shown in Fig. 1.

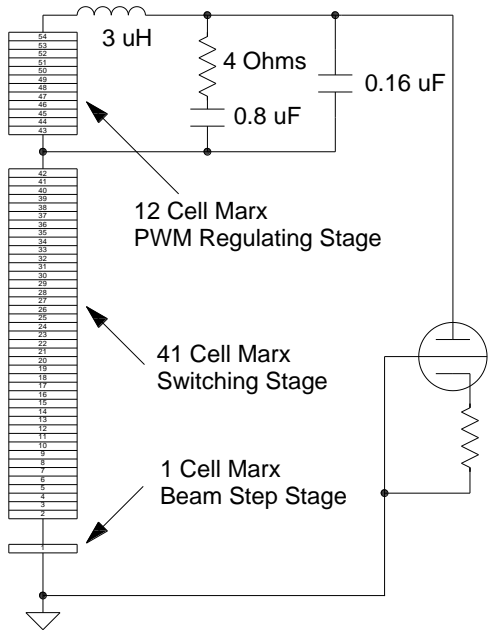

Figure 1: Full Marx modulator block diagram.

A simulation of the modulator is displayed in Fig. 2. The switching Marx cells are used to create the rising and falling edges and the regulating cells are used to do waveform shaping, after being filtered by a low pass filter. The switching cells are held constant during flattop to ensure no large transitions, except during the beam step. This step is required to compensate for accelerating cavity beam loading. These cells are added up together in series to create the output waveform. In this example, the capacitive droop is cancelled by the positive slope on the start of the flattop of the regulating cells, and the slope in the beam step portion is created with the negative slope in the regulating stage. This slope can be adjusted either by feedforward or feedback to compensate for both effects.

Furthermore, the beam step, which depends on the beam current being accelerated, needs to be settable anywhere from 0 to 10 kV. To accomplish that, a combination of the switching and special cells, powered by an adjustable power supply, are fired together to create the precise voltage step of any level.

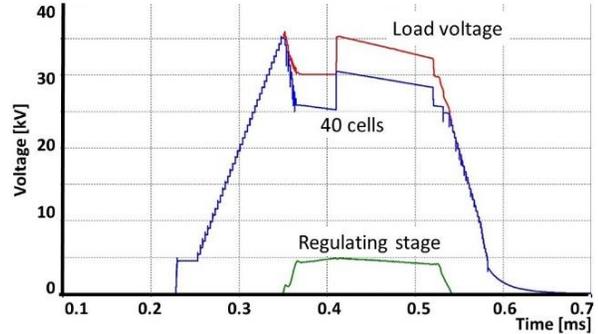

Figure 2: Modulator voltage waveform simulation.

## Switching Cells Design

The Marx-topology modulator works by charging individual cells in parallel and discharging in series. This allows for output voltage to be multiple of the input voltage. The power is delivered to each cell via charging diodes and a filter network. One side of a 1.7 kV half-bridge insulated-gate bipolar transistors (IGBT) is used to charge each cell through the previous cell. The large storage capacitors are then applied to the load via the fire IGBT, the other side of the half-bridge switch. This is illustrated in Fig. 3, where the conventional current path is shown with blue arrows for charging the cells and with red arrows for firing the cells into the load.

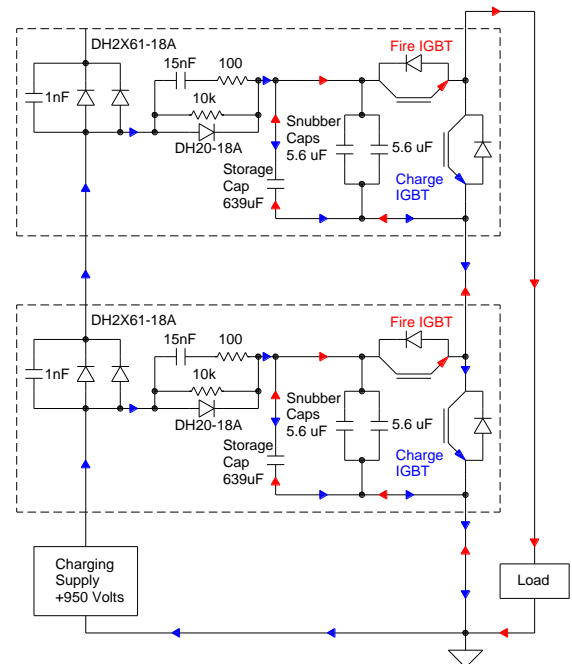

Figure 3: Marx double-cell circuit schematics.

In order to protect cells, redundant interlocks are used to insure the fire and charge IGBT's are not simultaneously on and each cell has dual 5.6 µF snubber capacitors to provide a low inductance path during high di/dt switching.

*PWM Regulating Cells*

Each regulating cell is fired via a pulse width modulation (PWM) scheme to create the waveform for compensating the capacitor droop, providing tilt on beam step and implementing limited voltage range real-time feedback. This is achieved by choosing one row of 12 regulating cells, after filtering, to do the fine waveform regulation. Although we desire to operate these cells during the entire pulse, they will only be activated for compensation during the flattop region of the waveform since the PWM cell switching losses prohibit operation longer than this time. To minimize the transition stress on each IGBT, a minimum ON and OFF time of 2 µs was selected. All of the cells will be interleaved every 1 µs, repeated every 12 µs, giving an 83.3 kHz repetition rate for each cell. This gives a minimum of 2 and maximum of 10 cells on at any time, giving a total of 8 cells, or 7.4 kV of adjustable range. The minimum and maximum pulse widths are illustrated in Fig. 4. The PWM switching cells, which are only active for 200 µs during the flattop region, give a junction temperature rise of less than 6 °C during each pulse, enabling long life for the IGBT modules. These cells are then filtered with a second order low pass passive network with a bandwidth of 100 kHz to reduce the ripple of the PWM regulating cells to less than ±25 V [4].

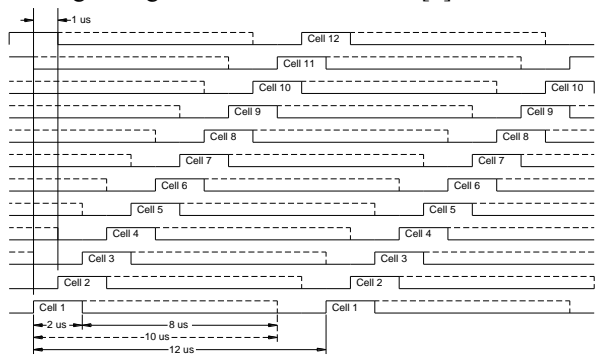

Figure 4: Regulating cells PWM waveforms.

The timing accuracy to set both the start time and width of the pulses is determined by the 100 MHz clock speed of the field programmable gate array (FPGA), an Altera Cyclone V SE SoC. To regulate both the switching and regulating cells, the FPGA will be used to program these cell timings with pulse to pulse learning. The PWM waveform will have the predominate ripple at 1 MHz, given by the interleaving, which will be filtered by the low-pass network on the output of the 12 regulating cells. It was also determined that the time delay between the fire command to the switching of the IGBT can vary, due to different fiber-optic cable length and turn on delay of each IGBT. According to simulations, greater than 20 ns of variation can cause ripple larger than the specification, which will be studied in more detail in the next prototype.

## EXPERIMENTAL RESULTS

The proposed Marx-topology modulator was first tested by constructing a single cell, followed by a 3, then a 9 cell version. As of the date of this publication, a 28 cell version is under construction. The cells have been corona tested up to 35 kV rms and have been tested into a shorted load at full current, achieving only 1.7 kA peak pulse for 10 µs. The 9 cell setup has also been tested for both single and multiple cell slew rate, with Fig. 5 showing the result of sequentially stepping up the cells. Finally the 9 cells were tested in PWM mode with a low pass filter on the output to ensure the ripple specification could be met. So far, the Marx-topology modulator has passed all tests during development. During the next stage, the 28 cells version will be tested with both switching cells, and PWM cells simultaneously before building the final full scale prototype of 54 cells. The results of these tests look promising for the final goal of an operating 35 kV modulator that meets all the required specifications.

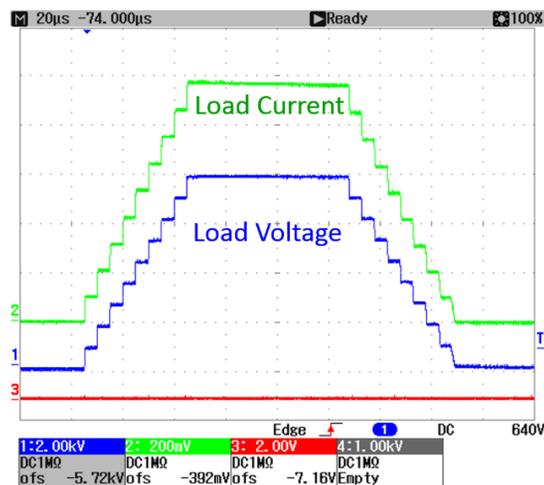

Figure 5: Waveforms for 9 Marx switching cells.
Ch. 2: load current (100A/V) and Ch. 1: load voltage

## ACKNOWLEDGMENT

The authors wish to acknowledge the contributions of J. Biggs, E. Claypool, N. Gurley, K. Martin, K. Roon, B. Stanzil, J. Walters, for their work in this development.

## REFERENCES

[1] F.G. Garcia et al., "Proton Improvement Plan Design Handbook", Fermilab Beams-doc-4053-v4, (2012).

[2] M.A. Kemp et al., "The SLAC P2 Marx", 2012 IEEE Power Modulator and High Voltage Conference, San Diego, CA, June 3-7, 2012, SLAC-PUB-15118.

[3] M.A. Kemp et al., "Design of a Marx-Topology Shaped-Pulse Modulator for FNAL," 2013 IEEE Pulsed Power and Plasma Science Conference, San Francisco, CA, USA, June 16-21, 2013.

[4] W.F. Praeg, "A High-Current Low-Pass Filter for Magnet Power Supplies," IEEE Transactions on Industrial Electronics and Control Instrumentation, Vol. IECI-17, No. 1, February 1970.